\documentclass[12pt]{article}
\usepackage{epsfig,graphpap,times}
\usepackage{amsmath}
\usepackage{amssymb}
\usepackage{amsthm,txfonts}
\usepackage[round,numbers,sort&compress]{natbib}
\usepackage{color}
\def\rbuildrel#1\over#2{\mathrel{\mathop{#2}\limits_{#1}}}

\catcode`@=11
\def\underline#1{\relax\ifmmode\@@underline#1\else
        $\@@underline{\hbox{#1}}$\relax\fi}

\def\titlepage{\pagestyle{empty}\c@page=0
      \def\thefootnote{\fnsymbol{footnote}} }
\def\endtitlepage{\pagestyle{plain}\c@page=1
      \def\thefootnote{\arabic{footnote}} \c@footnote\z@ }
\catcode`@=12

\newskip\humongous \humongous=0pt plus 1000pt minus 1000pt

\newif\ifdtup

\def\*{\hskip .06 cm}
\def\thebibliography#1{\section*{\ \markboth
 {REFERENCES}{REFERENCES}}\list
 {[\arabic{enumi}]}{\settowidth\labelwidth{[#1]}\leftmargin\labelwidth
 \advance\leftmargin\labelsep
 \usecounter{enumi}}
 \def\newblock{\hskip .11em plus .33em minus -.07em}
 \sloppy
 \sfcode`\.=1000\relax}

\newcommand{\beq}{\begin{equation}}
\newcommand{\enq}{\end{equation}}
\newcommand{\bqa}{\begin{eqnarray}}
\newcommand{\eqa}{\end{eqnarray}}
\newcommand{\bqas}{\begin{eqnarray*}}
\newcommand{\eqas}{\end{eqnarray*}}
\newcommand{\bea}{\begin{array}}
\newcommand{\ena}{\end{array}}

\date{}
\begin{document}

\begin{center}
\noindent{\large \bf Ion-mediated RNA structural collapse: effect of spatial confinement}\\
\end{center}
\vskip .1 in
\begin{center}
{\bf Zhi-Jie T{\footnotesize{\bf{AN}}}$^{1*}$  and
Shi-Jie C{\footnotesize{\bf{HEN}}}}$^{2*}$\\

$^1$Department of Physics and Key Laboratory of Artificial Micro \& Nano-structures of Ministry of Education,
School of Physics and Technology, Wuhan University, Wuhan, P.R. China 430072\\

$^2$Department of Physics and Astronomy and
Department of Biochemistry,
University of Missouri,
Columbia, MO 65211

\end{center}

\vskip .2 in
\centerline{\large \bf Abstract}
\vskip .1 in

\normalsize{ RNAs are negatively charged molecules residing in
macromolecular crowding cellular environments. Macromolecular
confinement can influence the ion effects in RNA folding. In this
work, using the recently developed tightly bound ion model for ion
fluctuation and correlation, we investigate the confinement effect
on the ion-mediated RNA structural collapse for a simple model
system. We found that, for both Na$^+$ and Mg$^{2+}$, ion
efficiencies in mediating structural collapse/folding are
significantly enhanced by the structural confinement. Such an
enhancement in the ion efficiency is attributed to the decreased
electrostatic free energy difference between the compact
conformation ensemble and the (restricted) extended conformation
ensemble due to the spatial restriction.
\\
}

\noindent -----------------------------------------------------------------------------------------------------------\\
\noindent{Running title: Ion-mediated RNA collapse and crowding effect}\\

\noindent{$^*$Emails: chenshi@missouri.edu (to S.J.C.); zjtan@whu.edu.cn (to Z.J.T.).}\\


\vfill\eject
\topmargin = -.2 in

\noindent{\large \bf Introduction}\\

Cellular functions of nucleic acids are intrinsically related to
their folding in cellular environments
\cite{Bloomfield,Tinoco,Bloomfield97}. Since DNAs/RNAs are highly
negatively charged molecules, folding into compact native
structures requires metal ions to overcome the strong
Coulombic repulsions \cite{Bloomfield}-\cite{Tan11}.
Furthermore,
RNAs {\it in vivo} are surrounded by many other
molecules \cite{Minton00,Zhou08,Burz09}, and the volume percentage of macromolecules in cells
can reach 40\% \cite{Minton00,Zhou08,Burz09}. The presence of other macromolecules
around an RNA can cause spatial confinement for DNA/RNAs folding.
The macromolecular crowding may greatly influence
the RNA folding
\cite{Draper10,Thirumalai08,Woodson10,Tanzheng10,Sugimoto10}.
However, the problem of
how the spatial constraint affects the ion-induced RNA folding
is not fully solved.
In this work, we will construct a simplified structural model to investigate the
macromolecular crowding effect on ion-mediated RNA collapse/folding.

For RNA secondary structure folding
\cite{Bloomfield,Tinoco,Bloomfield97}, the thermodynamic
parameters at standard salt condition (namely, 1M Na$^+$), have
been experimentally determined. Parameters for other ionic
conditions have also been fitted from experimental data and
theoretical calculations \cite{Serra95}-\cite{Tan2}. These
parameters have led to many accurate predictions for RNA/DNA
secondary structures and folding kinetics
\cite{Serra95}-\cite{Tan2}. For tertiary structure folding,
however, our understanding of the role of ions, especially the
effects of multivalent ions such as Mg$^{2+}$, remains incomplete
\cite{Bloomfield}-\cite{Tan11}. Experimental and computational
studies showed that metal ions with higher charge density are more
efficient in promoting RNA collapse and stronger ion-correlations
could potentially contribute to the efficient role of Mg$^{2+}$
ions \cite{Woodson1}-\cite{Lipfert2010}. Another series of
experiments for short DNA helices suggested the existence of
helix-helix attractive force at high multivalent ion concentration
and repulsive force at low ion concentration \cite{Qiu07,Qiu08}.
Furthermore, experiments on a paradigm system consisting of two
helices tethered by a loop suggested that Mg$^{2+}$ is $>$10-fold
more efficient than the predictions from the Poisson-Boltzmann
(PB) theory in promoting structural collapse
\cite{Bai1,Bai2,Chu09}.

In the cells, the presence of other molecules can cause an crowded
environment that can cause spatial confinement for the
conformations of RNAs and consequently influence the free energy
landscape of RNAs. For proteins, macromolecular crowding has been
shown to play an important role in folding stability, structure,
and kinetics
\cite{Minton00,Zhou08,Thirumalai05,Thirumalai09,Wang09,Zhou09,Liang10,Cheung11}.
For RNAs, however, the study of macromolecular crowding effect on
folding is relatively new \cite{Thirumalai11}, especially on its
impact on the role of ions in folding \cite{Woodson10}.

RNA tertiary structure collapse involves high charge
density, quantitative description for the role of metal
ions is challenging, especially for multivalent ions.
Molecular dynamics studies for ion-RNA interactions
have provided novel insights into
ion and structure/sequence-specific
forces in RNA folding, especially the force arising from
site-specific binding of ions \cite{Pappu09a}-\cite{Elber12}.
The two classic polyelectrolyte theories: the
counterion condensation (CC) theory
\cite{Manning1} and the PB
theory \cite{Gilson}-\cite{Lu} have been successful in predicting
electrostatic properties of biomolecules \cite{Manning1}-\cite{Lu}.
However, for the complex RNA
tertiary structures, the line-charge structural
model used in the CC theory could be
oversimplified \cite{Manning1}.
Furthermore, RNAs with high charge density
generally induce high ion concentration and consequently could possibly
cause ion-ion correlations
for multivalent ions in the vicinity of RNA surface. The PB
theory ignores such potentially important
effects for multivalent ion solutions \cite{Tan09,Tan1,Biopolymers}.
In order to take into account the effects of ion
correlations and ion-binding ensemble, the tightly bound ion (TBI)
model was developed \cite{Tan09,Tan1}; see the Supplementary Material
for an introduction of the TBI model. Experimental
comparisons suggest that
the TBI model may offer improved predictions on the
ion effect of thermodynamic stability of DNA/RNA helices/hairpins
\cite{Tan2,Tan5,Tan7}, DNA helix assembly
\cite{Tan3,Tan4,Tan6}, and ion-binding properties and tertiary folding
stability of RNAs in the presence of Mg$^{2+}$ \cite{Tan8,Tan9,Chen10}.

In this work, we will employ the newly refined TBI model
\cite{Tan8,Tan9} to investigate a paradigm system with two nucleic
acid helices tethered by a loop. We will focus on the effect of
conformational constraint on the role of ions in RNA folding. The
conformational constraint can arise from the macromolecular
crowding effect. Moreover, the study goes far beyond the previous
studies which only involves planar configurations \cite{Tan4}.
Specifically, we will consider the conformational fluctuation of
helices at the all-atom level with 3-dimensional rotations and
compare our predictions with the recent experimental data as well
as the predictions from the PB theory \cite{Bai2},
\\

\noindent{\large \bf Methods}\\

\noindent{\bf Structural model} \\

The model system consists of two helices tethered by a flexible
loop; see Fig. 1. In the model, we allow for three-dimensional
rotation of the helix axes. We note that in a previous study, only
symmetric co-planar model rotations were allowed \cite{Tan4}. As
shown in Fig. 1, a configuration of the system can be constructed
through the following operations. First, the two parallel helices
are separated by a axis-axis distance $x$; Second, the two helices
are symmetrically rotated around the respective ends (O and O') of
the helix axis in the axis-axis plane with an angle $\theta$;
Third, the two helices are rotated so that the projections of two
helix axes in the plane perpendicular to line of O-O' are
separated by an angle $\gamma$. The configuration of the system
can be described by three parameters: the distance $x$, the angles
$\theta$ and $\gamma$. Here, the parameter $\gamma$ is used to
produce non-planar configurations. The (DNA) helices here are
assumed to adopt the B-form structure \cite{Olson}. We use the
intervals of (3{\AA}, 20$^{\circ}$, 20$^{\circ}$) to sample the
conformational space of ($x \in [5\AA,56\AA]$, $\theta \in
[0,180^{\circ}]$, $\gamma \in [0,180^{\circ}]$). To reduce the
computational time, we randomly select $\sim$600 configurations
from the full 3D conformation ensemble for calculations. Tests
with the different sampling show that our results are quite
robust. In the structural model, the three
parameters (x, $\theta$, $\gamma$) are used to produce the major
portion of the full conformational ensemble, 
including the non-planar/planar and 
compact/extended conformational ensembles. 
Although this approximation may be valid as we 
focus on the low-resolution energy landscape 
of the system, 
it would be useful in further studies 
to examine the effects of 
other degrees of freedoms such as 
the spin of helices and 
asymmetrical rotations of helix axes.
\\

\noindent{\bf Free energy} \\

For a given configuration of the system described by
($x$,$\theta$,$\gamma$), the total free energy $G(x,\theta,\gamma)$ can
be calculated as
\beq
G(x,\theta,\gamma)=G_{E}(x,\theta,\gamma)+G_{\rm loop}(x,L)+G_{\rm
cs}+G_{\rm 3^\circ},\label{gland}
\enq
where
$G_E(x,\theta,\gamma)$ is the electrostatic free energy for the
two helices in an ionic solution.
$G_{\rm loop}(x,L)$ is the
free energy of the loop. $G_{\rm cs}$ is the coaxial stacking free energy
between the two helices, and $G_{\rm 3^{\circ}}$ is the
tertiary contact energy as the two helices are in close contact.

In the study, we calculate $G_E(x,\theta,\gamma)$ using the
all-atom TBI model \cite{Tan8,Tan9}; see the Supplemental
Information for a detailed description of the model. We calculate
$G_{\rm loop}(x,L)$ from \beq G_{\rm loop}=-k_BT {\rm  ln} P_{\rm
loop}(x,L), \label{gloop} \enq where $P_{\rm loop}(x,L)$ is the
probability of a loop of length $L$ with end-to-end distance $x$.
$P_{\rm loop}(x,L)$ can be estimated from an approximate
analytical expression \cite{Thirumalai}: $P_{\rm
loop}(x,L)=\frac{4\pi A (x/L)^2}{(1-(x/L)^2)^{9/2}}{\rm exp}\left(
\frac{-3L/l_p}{4(1-(x/L)^2)} \right)$, where $A$ is the
normalization constant \cite{Thirumalai,Thirumalai06,Hat} and
$l_p$ is the persistence length of the loop. In our calculations,
in order to make direct comparison with the experiment, we assume
that the loop is a PEG chain \cite{Bai2} and the persistence
length is 3.8{\AA} \cite{Tan4,Bai2}.

The coaxial stacking free energy term
$G_{\rm cs}$ is an important energetic factor
for nucleic acid helix alignment \cite{Walter94,Mathews07}. We
model this term based on the thermodynamic parameters \cite
{Walter94,SantaLucia98} and the RNA structure analysis \cite{Mathews07}
\bqa
G_{\rm cs}&=&G_0 \,\, {\rm for} \,\, x<10{\rm \AA} \,\, {\rm and} \,\, \theta>150^{\circ};\\
&=&0 \,\,\,\,\,\, \rm {else},
\eqa
where $G_0=G_{\rm \frac{XX}{XX}}-\delta g$.
Here, $\rm \frac{XX}{XX}$ is the coaxial
stacked base pairs, and $G_{\rm \frac{XX}{XX}}=-1.5$ kcal/mol for
$\rm \frac{CT}{GA}$ (at 25$^{\circ}$C) in the experimental system
\cite{Bai2}. $\delta g (\simeq -1$ kcal/mol) is an offset between
the coaxial stacking energy and the corresponding canonical base
pairs \cite{Walter94}.

$G_{\rm 3^\circ}$ is added to account for the possible tertiary contacts
\cite{Chu09}
\bqa G_{\rm 3^\circ}&=&g_{\rm 3^\circ} \,\, {\rm for} \,\, R_{\rm cc} \le 28{\rm \AA};\\
&=&0 \,\,\,\,\,\, \rm {else}, \label{g3} \eqa where $g_{\rm
3^\circ}$ is a constant negative energy. In the calculation, we
use -10$k_B$T, -12$k_B$T, and -14$k_B$T to investigate the effect
of the tertiary contact, respectively. We use the condition
$R_{\rm cc}\le 28{\rm \AA}$, a typical
inter-axial distance for ion-induced DNA aggregates
\cite{Rau92,Raspaud05,Todd08,Timsit10}, to simulate a close
contact where tertiary contact can occur. Our control tests show
that the small changes in the criteria around 28${\rm \AA}$ would
not cause significant changes in our quantitative predictions or
qualitative conclusions.

For the purpose of
visualizing the free energy landscape, we
will also use the inter-axis angle $\Theta$ to
represent the two angles $\theta$ and $\gamma$ (shown in Fig. 1)
\beq
\Theta ={\rm arccos}\left({\rm
cos}^2(\frac{\theta}{2}){\rm cos}^2(\frac{\gamma}{2})\right),
 \label{Theta}
\enq to show the electrostatic free energy
landscape in 2D ($x,\Theta$) plane instead of in 3D
($x,\theta,\gamma$) space.
\\

\noindent{\bf Spatial confinement} \\

As shown in Fig. 1,
we use $R_{\rm cc}$, the distance between the centers of the two axes of the helices,
to describe the compactness of the system.
Furthermore, we use the distance between the outer axial ends $R_{\rm PP'}\le R_{\rm max}$, to
quantify the spatial constraint of the system.
A smaller $R_{\rm max}$
corresponds to a stronger spatial confinement due to, for example,
macromolecular crowders.

To quantify the compactness of the system for an ensemble of
conformations, we use a Boltzmann-weighted averaged value
$\overline{R}_{\rm cc}$ for the distance $R_{\rm cc}$ between the
centers of the helix pair \cite{Tan4} \beq \overline{R}_{\rm
cc}=\frac{\sum_{(x,\theta,\gamma)}R_{\rm cc} {\rm
e}^{-G(x,\theta,\gamma)/k_BT}}{\sum_{(x,\theta,\gamma)}{\rm
e}^{-G(x,\theta,\gamma)/k_BT}}. \label{rg} \enq

\noindent{\large \bf Results and Discussions}\\

To investigate how the different ionic conditions affect the
folding properties of the model system, we apply the TBI model to
calculate the electrostatic free energy landscape for a wide range
of Na$^+$ and Mg$^{2+}$ conditions: [Mg$^{2+}$] $\in$ [0.01mM,
0.1M] and [Na$^+$] $\in$ [0M, 2M]. To directly simulate the
experiments \cite{Bai2}, we assume that the system is always
immersed in a 16mM Na$^+$ background (from the Na-MOPS buffer).
For convenience, we define the following two
specific states for the model system: (i) the random relaxation
state where the two helices can fluctuate randomly, corresponding
to the state with fully neutralized helices \cite{Bai1,Bai2}; (ii)
the (electrostatically) folded (collapsed) state with
$\overline{R}_{cc}\le 30 {\rm \AA}$
\cite{Bloomfield97,Tan3,Rau92,Raspaud05,Todd08,Timsit10}.

In the following, we will first present the results on the
electrostatic free energy landscape. We will then discuss the role
of ions in the structural collapse (without the possible tertiary
contact). Finally, we will discuss the effects of spatial
confinement and tertiary contacts on the ion role in structural
folding.
\\

\noindent {\bf Electrostatic free energy landscape}\\

\noindent {\it In Na$^+$ solutions.}

Figs. 2A-D show the electrostatic free energy landscape $G_E$ for
the different added [Na$^+$]. At low added [Na$^+$], the helices
tend to avoid each other with the largest $x$ and $\Theta$ (see
Fig. 2A) due to the weak charge neutralization. An increased
[Na$^+$] would enhance the ion binding and charge neutralization
due to a lower entropic penalty for ion-binding, thus causing a
weakened  inter-helix repulsion. As shown in Fig. 2B, the helices
fluctuate around less extended conformations with smaller $x$ and
$\Theta$ values. When [Na$^+$] becomes very high ($>0.3$M), the
helices become nearly fully neutralized, thus the helices can
fluctuate randomly in the full conformational space (see Figs.
2CD), except for very compact configurations (very small $x$ and
$\Theta$) where helix-helix and ion-helix exclusions play a
dominate role.
\\

\noindent {\it In Mg$^{2+}$ solutions.}

Figs. 2E-H show the electrostatic free energy landscapes $G_E$ for
the helices with the different added [Mg$^{2+}$]. At low
[Mg$^{2+}$], the helices form extended conformations with large
$x$ and $\Theta$. With the increase of [Mg$^{2+}$], the two
helices switch to the conformations with smaller $\Theta$ and $x$
(Figs. 2FG). For high [Mg$^{2+}$] ($> 3$ mM), the helices tend to
adopt the conformations with very small $\Theta$ ($\Theta\lesssim
40^{\circ}$) and intermediate $x$ ($32{\rm \AA} \lesssim x
\lesssim 47 {\rm \AA}$); see Fig. 2H. The above transition with
increasing [Mg$^{2+}$] can be attributed to the divalent
ion-mediated helix-helix interactions \cite{Tan3,Tan4,Bai1}. When
[Mg$^{2+}$] is very low, background Na$^+$ of low concentration in
buffer only gives weak neutralization to helices, thus the helices
repel each other and the conformations with largest $x$ and
$\Theta$ would be favorable. With the increase of [Mg$^{2+}$],
Mg$^{2+}$ will compete with background Na$^+$, and the
self-organization of Mg$^{2+}$ could induce (slight) attractive
force between helices \cite{Tan3,Tan4}. Such attraction would
become stronger for the helix-helix near-parallel configurations
at intermediate separation and at higher [Mg$^{2+}$] \cite{Tan4}.
As a result, the helices would transform to the conformations with
small $\Theta$ and intermediate $x$ \cite{Tan3}.

Compared with the results
for 2D (co-planar) configurations (Fig. 2 in Ref \cite{Tan4}),
more extended conformations (large $\Theta$) are shown in the
current 3D free energy landscapes, causing an increase in the conformational
entropy of the two helices and weakening the effect of
the possible Mg$^{2+}$-mediated helix-helix attractive force
in the structural collapse.
\\

\noindent {\bf Ion-mediated structural collapse: without spatial confinement}\\

First, we discuss the role of ions in the structural collapse without
spatial confinement and
tertiary contact by setting $G_{\rm 3^\circ}=0$ in Eq. \ref{gland}.
In the following, we present general features for the system and the comparisons with
the experimental data \cite{Bai2}.  \\

\noindent {\it General features}

As shown in Fig. 3A, with the increase of [Na$^+$] to 2M,
$\overline{R}_{\rm cc}$ decreases from a large value to that of
the random relaxation (fully neutralized) state. Such a trend of
$\overline{R}_{\rm cc}$ is a results of the electrostatic free
energy landscapes shown in Fig. 2A-D and can be attributed to the
aforementioned enhanced Na$^+$-binding at higher [Na$^+$]. Also
shown in Fig. 3A is the loop length dependence of the structural
compaction at low and high [Na$^+$]. A PEG chain is quite flexible
($l_p\simeq 3.8 {\rm \AA}$). The loop (length $L \gg l_p$) tends
to minimize the loop entropic loss upon structure formation.
Therefore, the loop drives the system to fluctuate around a
relatively smaller end-to-end distance $x$ since a larger $x$
corresponds to a smaller loop entropy. For example, for PEG loops
with $L=56 {\rm \AA}$ and $32 {\rm \AA}$, $x\simeq 20 {\rm \AA}$
and $15 {\rm \AA}$, are the most favorable according to Eq.
\ref{gloop}, respectively. The loop effect is slightly stronger
for low [Na$^+$] where the favorable configurations are extended
and the effect of the (short) tether can become more important.

In a solution with added [Mg$^{2+}$], at low [Mg$^{2+}$], the
favorable conformations are those with large $x$ and $\Theta$,
thus a loop length can contribute to the compaction of the system.
In contrast, at a high [Mg$^{2+}$], Mg$^{2+}$ induces (slightly)
attractive force between helices at small inter-axis angles
$\Theta$ (see Fig. 2H). When the helices are driven by the loop to
a smaller $x$ (to gain loop conformation entropy), the low free
energy conformations (in the small $x$ regime in Figs. 2GH)
correspond to a range of $\Theta$. As a result, a small $x$ does
not necessarily lead to compaction of the system (characterized by
a small $\overline{R}_{\rm cc}$). We note that a high [Mg$^{2+}$]
can induce a state that is slightly more compact than the random
relaxation (fully neutralized) state yet much looser than the
upper limit of the electrostatically folded state
\cite{Bloomfield97,Rau92,Raspaud05,Todd08,Timsit10}, as shown in
Fig. 3B.
\\

\noindent {\it Comparisons with experiments}

To make direct comparison with the experimental system
\cite{Bai2}, we assume the loop length to be $L=32 {\rm \AA}$. We
calculated the Boltzmann-weighted averaged SAXS profiles $I(Q)$
over conformation ensemble for each ionic condition. Here $Q$ is
the scattering vector which is equal to $4\pi sin({\it
\vartheta})/\lambda$, where ${\vartheta}$ is the Bragg angle and
$\lambda$ is the wave length of the radiation.

Figs. 3C\&D show the SAXS profiles for the different added [Na$^+$]
and [Mg$^{2+}$], respectively.
As the added [Na$^+$] is increased from 0 to 2M
or Mg$^{2+}$ is added from 0 to 0.1M,
the SAXS profile changes from that of the extended state to that
of the nearly fully neutralized state.
In addition, the profiles indicate that the system
at high [Mg$^{2+}$] ($\sim$ 0.1M) is slightly more compact than
that at high [Na$^+$] ($\sim$ 2M). Since the experimental SAXS
profiles are in arbitrary unit and with vertical shifts \cite{Bai2},
we cannot make direct comparisons for the SAXS profiles
between the experiments and our predictions.

Based on the calculated SAXS profiles and compactness
$\overline{R}_{\rm cc}$, we can estimate the fraction ``relaxed''
$\chi$ for the system by \cite{Bai2} \beq {\rm \Omega\, (Na^+ \,
or \, Mg^{2+})=(1-\chi)\, \Omega\, (no \, added \, salt)+\chi\,
\Omega\, (high \, salt)}, \label{xi} \enq where $\Omega$ stands
for a structural parameter, such as $I(Q)$ or $\overline{R}_{\rm
cc}$. ${\rm \Omega\, (no \, added \, salt)}$ is that in 16mM
Na$^+$ background from the Na-MOPS buffer without added salt, and
${\rm \Omega (high \, salt)}$ is taken as that of the
fully-neutralized state in our estimation. Based on the equation,
we estimate $\chi$ from the calculated $I(Q)$ and
$\overline{R}_{\rm cc}$ respectively, and compare our predictions
with experiments for the cases with added Na$^+$ and Mg$^{2+}$.

Figs. 3E\&F show the fraction ``relaxed'' $\chi$ as a function of
added [Na$^+$] and [Mg$^{2+}$], respectively. When [Na$^+$] is
added, $\chi$ increases from zero to $\sim$ 0.9 at 2M [Na$^+$],
suggesting that the system is close to the random relaxation
(fully neutralized) state. The lines estimated from $\it I(Q)$ and
$\overline{R}_{\rm cc}$ are almost identical, and the predictions
agree well with the experimental data, as shown in Fig. 3E.

As shown in Fig. 3F, when Mg$^{2+}$ is added, $\chi$ increases
from zero to $\sim$ 0.9 at $\sim$8mM [Mg$^{2+}$], and may slightly
exceeds 1 when [Mg$^{2+}] \gtrsim$ 10mM, suggesting slightly more
compact state induced by high [Mg$^{2+}$]. In addition, the
predicted $\chi$ from $\overline{R}_{\rm cc}$ is slightly larger
than that from $\it I(Q)$ at high [Mg$^{2+}$] (shown in Fig. 3F),
which is in accordance with the above described relations of
$\overline{R}_{\rm cc}$ and $\it I(Q)$ with [Na$^+$] and
[Mg$^{2+}$]. Fig. 3F also shows that the predicted $\chi$'s from
the TBI model are in good accordance with the experimental data,
while PB theory predicts $>$10-fold higher midpoint of [Mg$^{2+}$]
for the Mg$^{2+}$-mediated collapse. This result may be attributed
to the ignored ion-ion correlations in the PB. The inclusion of
ion correlations would allow ions to self-organize to form
low-energy state, and thus the TBI model predicts a more efficient
role of Mg$^{2+}$ than the PB \cite{Tan1,Tan8,Tan9}.

It is important to note that
although the system has the similar global compactness
at high [Na$^+$] and [Mg$^{2+}$], the free energy landscapes are
very different as shown in Fig. 2.

The PEG loop used in the above calculation is electrically
neutral, while a realistic RNA loop is a
polyanionic chain. Recent experiments suggest that the hinge
stiffness can become an energetic barrier for RNA folding
\cite{Pollack2}. As shown previously \cite{Tan4}, a nucleotide
loop may enhance the sharpness of ion-mediated structural collapse
due to the ion-dependence of the persistence length of the loop
(e.g., \cite{Tan4}).
\\

\noindent{\bf Ion-mediated folding: with spatial confinement}\\

In this section, we investigate the influence of spatial
restriction as well as a tertiary contact on the ion-mediated
collapse of the above two-helix system.
\\

\noindent{\it Electrostatic free energy landscape with spatial confinement.}

With the conformational confinement defined by the limit $R_{\rm
max}$ for the distance $R_{\rm PP'}$ between the outer ends of the
two helices (see Fig. 1), the conformational space of the system
is confined. The electrostatic free energy landscape of the
confined system $G_E(R_{\rm cc},R_{\rm max})$ is given by \beq
G_E(R_{\rm cc},R_{\rm max})= -k_BT \ln \sum_{R_{\rm cc}}^{R_{\rm
PP'}\le R_{\rm max}} e^{-G_E(x,\theta,\gamma)/k_BT}, \label{Ger}
\enq where $R_{\rm cc}$ is the distance between the two centers of
the two helices (see Fig. 1).

As shown in Fig. 4, the spatial restriction influences the free energy landscape
significantly for both Na$^+$ and Mg$^{2+}$ solutions.  For large
$R_{\rm max}$, the free energy landscape is close to that without spatial restriction.
At low [Na$^+$], the favorable conformations are clustered in the region of
large $R_{\rm cc}$. At high [Na$^+$] (e.g., 2M), the helices can fluctuate
nearly in the whole region with all the different $R_{\rm cc}$ values.  With
the decrease of $R_{\rm max}$, the conformations of the two helices are
confined by the restriction and the extended conformations are severely
limited.  This causes the free energy $G_E(R_{\rm cc},R_{\rm max})$ at high
[Na$^+$] to become close to that at low [Na$^+$].  In the limit of strong
spatial confinement such as $R_{\rm max}\le 40\AA$, the free energy
$G_E(R_{\rm cc},R_{\rm max})$ at the different Na$^+$ concentrations can become nearly
identical; see Figs. 4A-D in the small $R_{\rm max}$ regime.

For Mg$^{2+}$ solutions, the spatial confinement
effect on $G_E(R_{\rm cc},R_{\rm max})$ is similar to that for Na$^+$ solutions,
except for
the much more efficient role of
the Mg$^{2+}$ ions in promoting the collapse of the system.
The electrostatic free energy landscapes show that
with the spatial confinement,
the extended conformations are much more
significantly restricted than the compact conformations,
causing the destabilization of the extended state.
In the ion-promoted collapse, the spatial confinement
would effectively enhance the ion effect in folding.
\\

\noindent{\it Ion-mediated folding: effects of spatial confinement and tertiary contacts}

To investigate the role of tertiary contacts in ion-induced
structural collapse, we add a negative constant energy to model
the effect of the tertiary contacts. Specifically, we choose three
values -10$k_B$T, -12$k_B$T, and -14$k_B$T for $g_{\rm 3^\circ}$
(see Eq. \ref{g3}).

Helix-helix attraction arising from the tertiary contacts would
cause a shift in the conformational distribution toward the
compact state and lower the ion concentration required to induce
the collapse of the structure. Furthermore, the tertiary contact
results in a much sharper ion-induced structural
collapse (than the tertiary contact-free case). The effect is
more pronounced for stronger tertiary contacts. As shown in Figs.
5ACE for Na$^+$, the midpoint for the folding transition occurs at
[Na$^+$]$\sim 0.5$M for $g_{\rm 3^\circ}=-10k_BT$ and
[Na$^+$]$\sim0.08$M for $g_{\rm 3^\circ}=-14k_BT$, respectively,
and the stronger tertiary contacts cause a slightly more compact
state. As shown in Figs. 5BDF, the midpoints of of
Mg$^{2+}$-dependent folding are $\sim$0.8mM and $\sim$0.2mM for
$g_{\rm 3^\circ}=-10k_BT$ and $g_{\rm 3^\circ}=-14k_BT$,
respectively. Again, we note that much lower [Mg$^{2+}$] ($\sim$
mM) is required to cause folding than [Na$^+$] ($\sim$ M), which
is in accordance with the experimental data
\cite{Tan11,Woodson1,Tan8,Tan9}.

To further examine the effect of the spatial confinement in the
presence of tertiary contacts, we calculated $\overline{R}_{\rm
cc}$ with a given $R_{\rm max}$ constraint (the maximum distance
of P-P' shown in Fig. 1). As shown in Figs. 5A-F, the spatial
confinement significantly enhances the ion efficiency in mediating
the structural collapse/folding in both Na$^+$ and Mg$^{2+}$
solutions. For example, for $g_{\rm 3^\circ}=-10k_BT$, with the
decrease of $R_{\rm max}$ from 120${\AA}$ to 70${\AA}$, the
[Na$^+$] and [Mg$^{2+}$] at the midpoint of the folding transition
decreases from $\sim$0.5M to $\sim 0.1$M and from $\sim$1mM to
$\sim$0.3mM, respectively. For a stronger tertiary contact, e.g.,
$g_{\rm 3^\circ}=-14k_BT$, the spatial confinement would further
enhance the ion efficiency for both of Na$^+$ and Mg$^{2+}$, as
shown in Figs. 5A-F. Such higher efficiency of
Mg$^{2+}$ than Na$^+$ is attributed to the higher valence of
Mg$^{2+}$ which corresponds to the less entropy for ion binding
and stronger ion-ion correlation, causing higher Mg$^{2+}$ binding
affinity \cite{Tan8} and consequently higher efficiency in
stabilizing RNA \cite{Tan9}.

The above spatial confinement-promoted ion (Na$^+$ and Mg$^{2+}$)
efficiency in mediating structural folding is in accordance with
the recent experiment which showed that a lower Mg$^{2+}$ is
required to fold Azoarcus ribozyme in PEG environments
\cite{Woodson10}.  The spatial confinement can be caused by
molecular crowding or other {\it in vivo} effects.
The spatial confinement can destabilize the
low-salt (extended) state by reducing the conformational space and
consequently decrease the electrostatic free energy difference
between the compact conformations and the (restricted) extended
conformations \cite{Thirumalai11}. In addition, the tertiary
contacts would further stabilize the compact state. These two
effects combined together cause a reduced ion concentration
required to induce the folding transition.
Therefore, Na$^+$ and Mg$^{2+}$ are more efficient in RNA
folding in a crowded environment.  \\

\noindent{\large \bf Conclusions and Discussions}\\

In this work, we employed the newly refined TBI model
\cite{Tan8,Tan9} to investigate how ions cause
compaction for a system of loop-tethered helices.
The main advantage of the TBI model is its ability to
account for the possible ion correlation and fluctuation effects.
The present study distinguishes from the
previous studies \cite{Tan4} on similar systems in several novel aspects.

First, the present study is based on the
3D (instead of planar) rotational degrees of freedom
in the conformational sampling of the system.
In contrast to co-planar (2D) helix orientations,
the 3D random orientation of the helices would
reduce the (ensemble averaged) ion correlation effect and the associated force
between the helices.

Second, the main focus of the present study is the influence of
the spatial confinement and the tertiary contacts upon the ion
effect and the comparisons with the the PB-based predictions and
the recent experimental results. The study leads to the following
main conclusions:
\begin{enumerate}
\item Na$^+$ can induce a transition from an ensemble of extended
conformations to an ensemble of random relaxed conformations with
a flat energy landscape. In contrast, Mg$^{2+}$ can induce a
transition from an extended state to a conformation ensemble that
is slightly more ordered and slightly more compact than the random
relaxation conformational ensemble. However, such a
Mg$^{2+}$-induced state is far looser than the (electrostatically)
folded state. The predictions from the TBI model are in accordance
with the experimental data for both Na$^+$ and Mg$^{2+}$
solutions, while the PB theory significantly underestimates the
Mg$^{2+}$ efficiency.

\item A tertiary contact would make the ion (Na$^+$ and
Mg$^{2+}$)-dependent folding transition much sharper than the
tertiary contact-free case. Furthermore, stronger tertiary
contacts can fold RNAs at lower ion concentrations for both of
Na$^+$ and Mg$^{2+}$.

\item Spatial confinement (e.g., due to macromolecular crowding)
significantly promotes the ion efficiency in mediating RNA folding
for both Na$^+$ and Mg$^{2+}$. The effect of the spatial
confinement can be attributed to the decreased
electrostatic free energy difference between the compact ensemble
and the (restricted) extended ensemble.
\end{enumerate}

The current theory gives overall good agreement with the
experiment results, suggesting an overall reliable analysis and
predictions for the ion-induced compaction and the effects of
spatial confinement. Further investigation of the problem requires
improvements of the model in several aspects. First, in the
structural modeling, although we have taken in
account the non-planar conformations, we have ignored 
conformations generated from the spin of the helices 
and the nonsymmetric rotations of the helix axes.
Including these conformations
may cause changes in the free energy landscape, though 
a detailed study for such changes requires
exceedingly high computational cost.  
Second, the current form of the TBI
model cannot treat the sequence preference in ion-binding
\cite{Tan6,Stellwagen07}. The specific binding could make
important contributions to the ion effect, although for the
PEG-tethered helix system here, the contribution of the specific
binding may be weak. Finally, we have neglected the interference
from the helix on the loop conformations except for the constraint
of the end-end distance $x$. The effect of the loop-helix
interference may not be significant for a PEG loop. However, such
a effect can be important for a polynucleotide loop which is
highly charged.
\\

\noindent{\bf \large Acknowledgments}\\

This research was supported by the National Science Foundation
(grants MCB0920067 and MCB0920411), National Institutes of Health
(grant GM063732) (to S.-J.C.), National Science Foundation of
China (grants 11074191, 11175132, and 10844007), Program for New
Century Excellent Talents (NCET 08-0408), Fundamental Research
Funds for the Central Universities, the National Key Scientific
Program (973)-Nanoscience and Nanotechnology (No. 2011CB933600)
and the Scientific Research Foundation for the Returned Overseas
Chinese Scholars, State Education Ministry (to Z.-J.T.).

\newpage
\centering{\bf FIGURE CAPTION}
\begin{description}

\item [FIGURE 1]
The conformation of two (DNA)
helices tethered by a loop can be characterized by three
structural parameters: the end-to-end distance $x$ (O-O'), the angles
$\theta$ and $\gamma$ illustrated in the figure.
The freedom of $\gamma$ (rotation of the
axis into the paper) is used to produce non-planar configurations.
The P-P' distance $R_{\rm PP'}\le R_{\rm max}$ represents
the spatial restriction for the two helices
due to, e.g., the macromolecular crowding effect.
$R_{\rm cc}$ between the centers
of the helices is used to quantify the compactness of the model.
Tertiary contact energy can be added for conformations with a small $R_{\rm cc}$.
The DNA helices are constructed with the software X3DNA \cite{Olson}
and are displayed with the software PyMol
(http://pymol.sourceforge.net/).

\item [FIGURE 2] The electrostatic free energy landscapes
$G_E(x,\theta,\gamma)$ as a function of [Na$^+$] (A-D) and
[Mg$^{2+}$] (E-F). Here $x$ is the end-to-end distance and
$\Theta$ is the angle between the two helix axes. $\Theta$ is
determined by the ($\theta$,$\gamma$) angles (see Eq.
\ref{Theta}). The color scale represents the energy difference
between the configuration ($x$, $\Theta$) and the state with the
minimum free energy. From 0 to 5 $k_BT$, the color
changes gradually from Red to Blue.
Note that the system is in
a 16mM Na$^+$ background due to the Na-MOPS buffer \cite{Bai2}.

\item [FIGURE 3]
Ion-mediated structural collapse in the absence of the tertiary contact and
spatial confinement.
(A, B) Ion concentration-dependence of the compactness $\overline{R}_{\rm cc}$
(the mean distance between the helix centers) as a function of Na$^+$
(A) and Mg$^{2+}$ (B). Dashed lines denote the random relaxation
(fully-neutralized) state
and the dotted lines denote the upper limit
of the electrostatically folded state
($\overline{R}_{\rm cc}\sim 30{\rm \AA}$) \cite{Tan3,Tan4,Rau92,Raspaud05,Todd08}.
(C,D)
The SAXS profiles calculated for the different added Na$^+$ (C) and
Mg$^{2+}$ (D).
Lines are the reference SAXS profiles.
Dashed lines: the upper limit
of the electrostatically folded state ($\overline{R}_{\rm cc}\simeq 30 {\rm \AA}$);
Solid lines: the random relaxation (fully-neutralized) state;
Dotted lines: the extended state (with no added salts).
The SAXS profiles are calculated from the software
CRYSOL \cite{Koch} with the statistical weight determined from
the free energy $G(x,\theta,\gamma)$.
(E,F) The ``relaxed'' fraction $\chi$
as a function of [Na$^+$] (E) and [Mg$^{2+}$] (F), respectively; see Eq. \ref{xi}.
Symbols: the experimental data by Bai et al \cite{Bai2};
Solid lines: the calculated results with the SAXS profiles $I(Q)$
determined from the TBI model;
Dotted lines: calculated with
the compactness $\overline{R}_{\rm cc}$ determined from the TBI model;
Dashed lines: calculated with
the SAXS profiles determined from the PB theory by Bai et al \cite{Bai2}.
The ``relaxed''
state is taken as the fully neutralized state.
Note that the system is in
16mM Na$^+$ background due to the Na-MOPS buffer \cite{Bai2}.

\item [FIGURE 4] The electrostatic free energy landscapes
$G_E(R_{\rm cc}, R_{\rm max})$, where
$R_{\rm max}$ represents the spatial restriction for
the helices (e.g., due to the macromolecular crowding); see Eq. \ref{Ger}.
The color scale denotes the energy difference between
$G_E(R_{\rm cc}, R_{\rm max})$ and the minimum value for each $R_{\rm max}$:
$\Delta G_E=G_E(R_{\rm cc},R_{\rm max})-G_E(R_{\rm cc},R_{\rm max})_{\rm min}$.
From $0$ to $5 k_BT$, the color changes uniformly from Red to Blue.

\item [FIGURE 5]
Ion-mediated structural folding in the presence of tertiary contact and spatial restriction.
(A,B) The compactness $\overline{R}_{\rm cc}$ of the loop-tethered helices
as functions of [Na$^+$] (A) and [Mg$^{2+}$] (B)
for the different $R_{\rm max}$ values
(from the right to the left: 120$\AA$, 82$\AA$, 78$\AA$, and 70$\AA$)
and the different tertiary contact energy $g_{\rm 3^\circ}$
(solid lines: -10$k_BT$, and dashed lines: -14 $k_BT$).
(C,D) The folded fraction $f_{\rm folded}$  as a function of [Na$^+$] (C) and [Mg$^{2+}$] (D)
for the different $R_{\rm max}$ values
(from the right to the left: 120$\AA$, 82$\AA$, 78$\AA$  and 70$\AA$)
and the different tertiary contact energy $g_{\rm 3^\circ}$
(solid lines: -10$k_BT$, and dashed lines: -14 $k_BT$).
(E,F) The ion concentration midpoints for the folding transition
as a function of $R_{\rm max}$ and the different tertiary contact energies $g_{\rm 3^\circ}$.
From the top to the bottom: $g_{\rm 3^\circ}$=-10$k_BT$, -12$k_BT$, and -14$k_BT$, respectively.
\end{description}

\newpage

\begin{figure}[ht]
\begin{center}
\epsfxsize= 8.0cm \centerline{\epsfbox{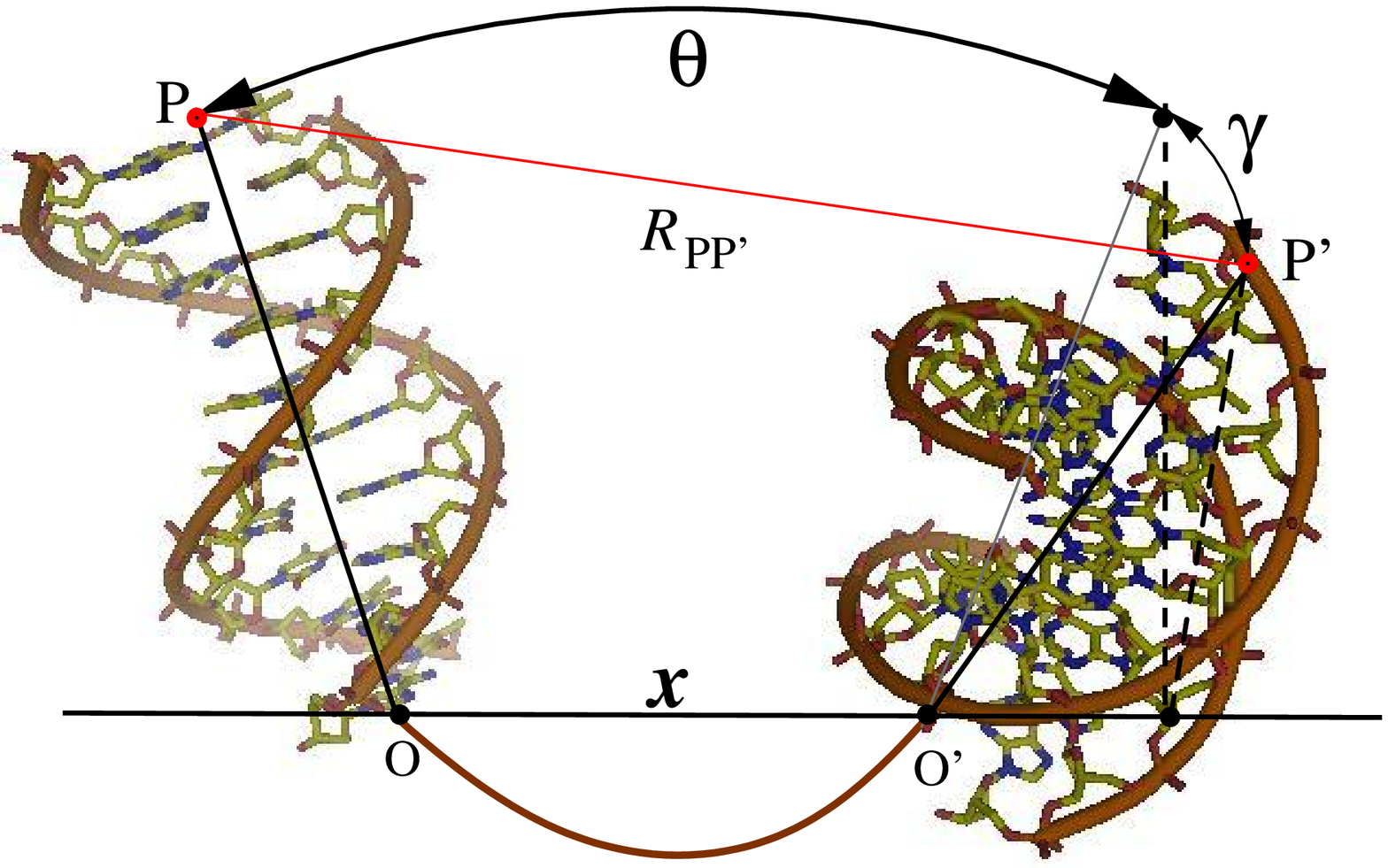}}
\end{center}
\caption{
} \label{setup}
\end{figure}

\newpage
\begin{figure}[ht]
\begin{center}
\epsfxsize= 16.0cm \centerline{\epsfbox{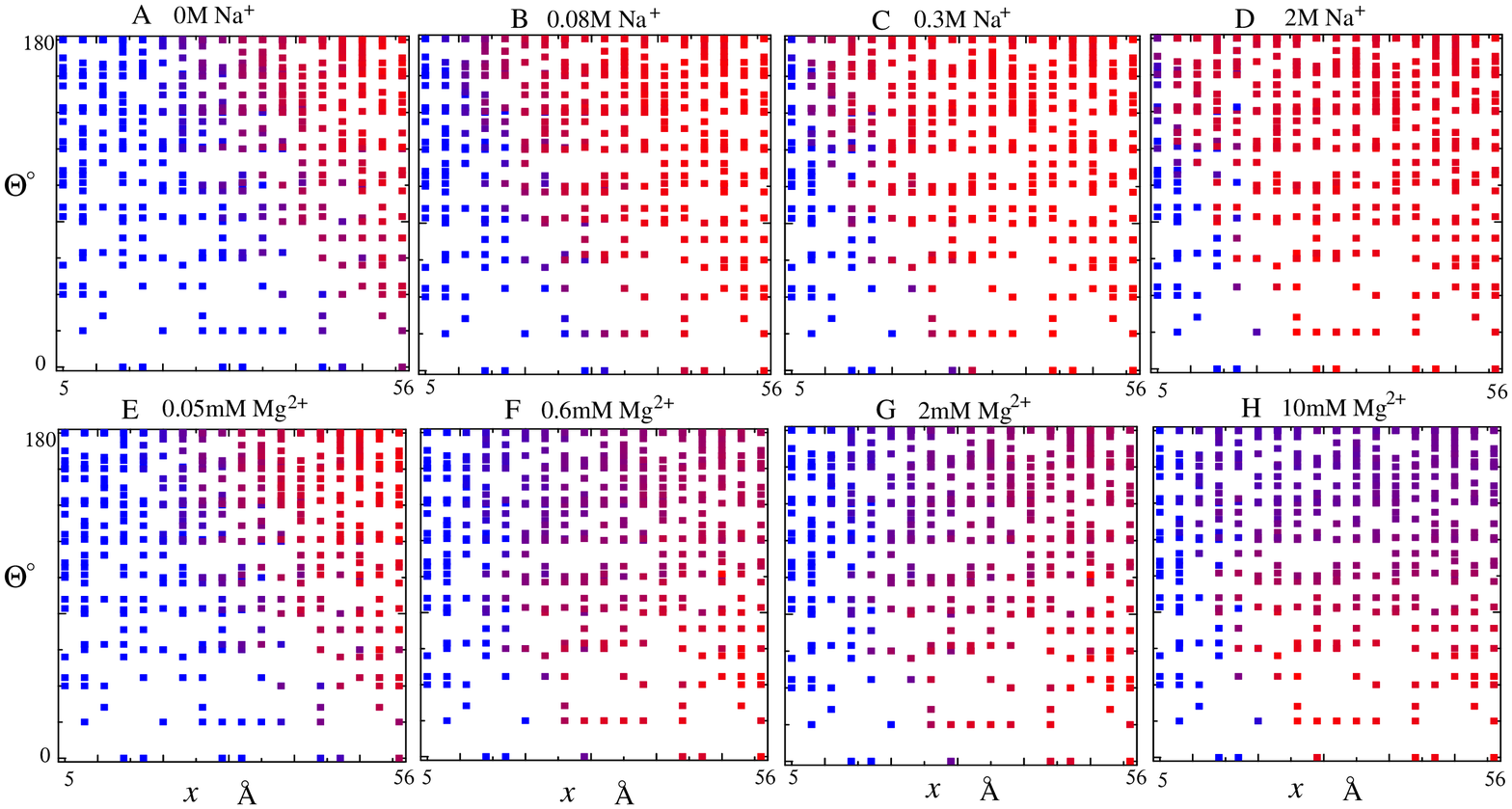}}
\end{center}
\caption{
} \label{setup}
\end{figure}

\newpage
\begin{figure}[ht]
\begin{center}
\epsfxsize= 16cm \centerline{\epsfbox{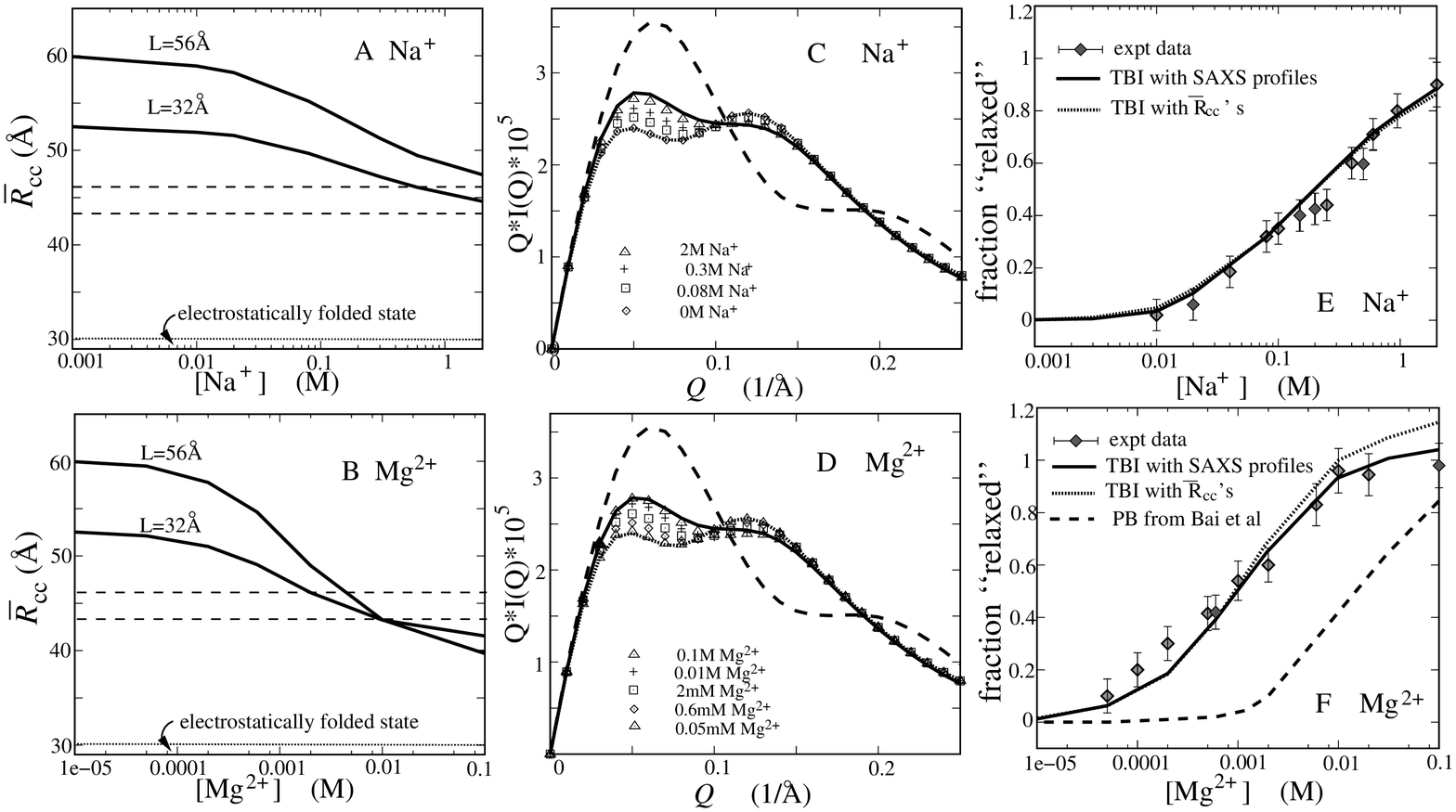}}
\end{center}
\caption{
} \label{setup}
\end{figure}

\newpage
\begin{figure}[ht]
\begin{center}
\epsfxsize= 16cm \centerline{\epsfbox{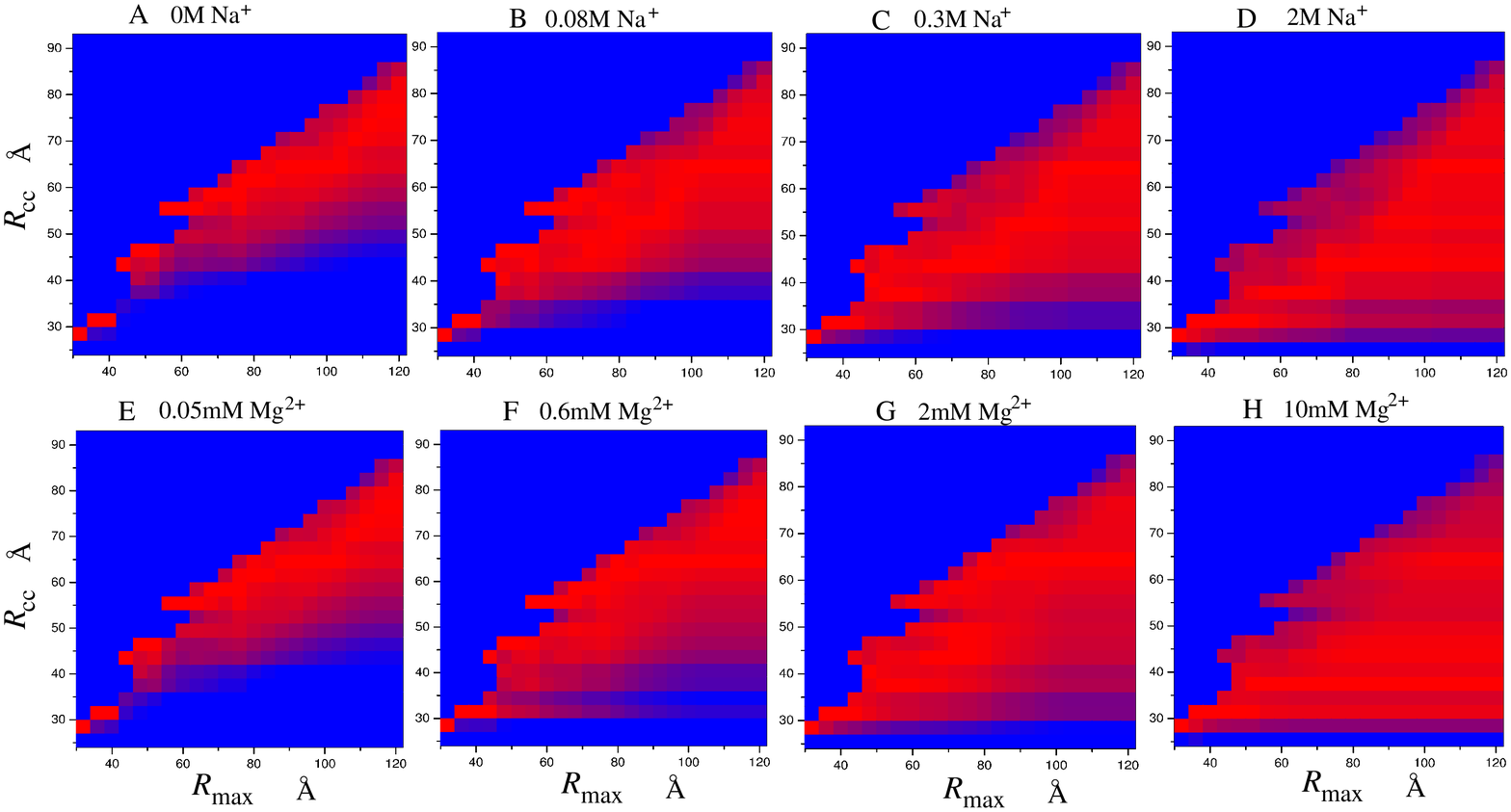}}
\end{center}
\caption{
} \label{setup}
\end{figure}

\newpage
\begin{figure}[ht]
\begin{center}
\epsfxsize= 16cm \centerline{\epsfbox{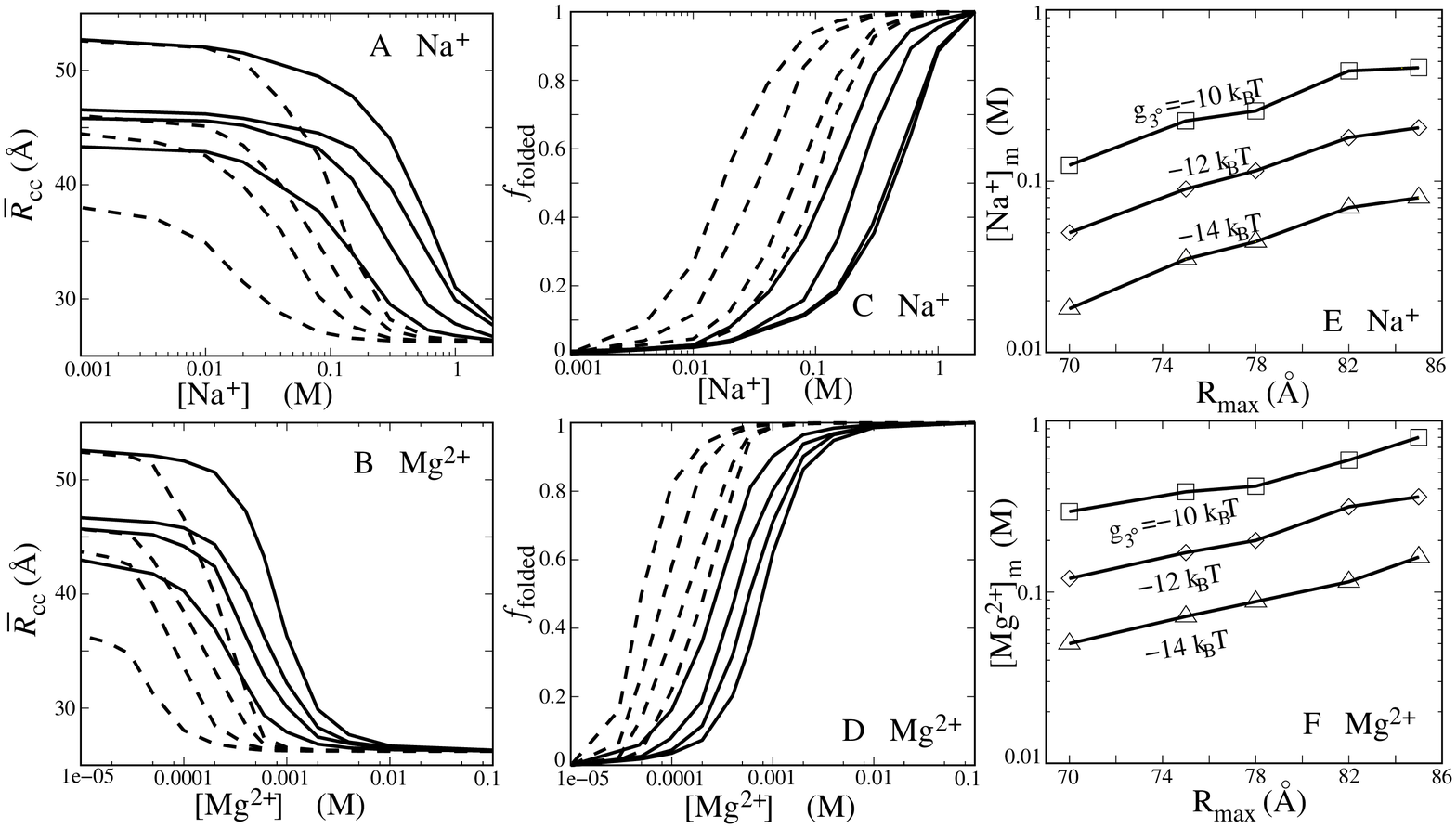}}
\end{center}
\caption{
} \label{setup}
\end{figure}

\end{document}